The Youthful Appearance of the 2003 $EL_{61}$ Collisional Family

David L. Rabinowitz[1], Bradley E. Schaefer[2], Martha Schaefer[3], Suzanne W. Tourtellotte[4]

[1] Center for Astronomy and Astrophysics, Yale University, P. O. Box 208121, New Haven CT 06520-8121 email: david.rabinowitz@yale.edu

[2] Department of Physics & Astronomy, Louisiana State University, 243 Nicholson, Baton Rouge LA 70803-0001

[3] Department of Geology & Geophysics, Louisiana State University, 341 Howe-Russell, Baton Rouge LA 70803

[4] Astronomy Department, Yale University, P. O. Box 208121, New Haven CT 06520-8121

ABSTRACT

We present new solar phase curve observations of the 2003 $EL_{61}$ collisional family showing that all the members have light-scattering properties similar to the bright icy satellites and dwarf planets. Compared to other Kuiper Belt objects, the five family members we observe (2003 $EL_{61}$, 2002 $TX_{300}$, 2003 $OP_{32}$, 2005 $RR_{43}$, and 1995 $SM_{55}$) have conspicuously neutral color (V-I = 0.6-0.8 mag) and flat phase curves at small phase angles (phase coefficients of 0.0 - 0.1 mag $deg^{-1}$). Comparing the phase curves we observe for other icy Kuiper Belt objects to the phase curves of icy satellites, we find that the flat phase curves of the 2003 $EL_{61}$ family are an indication they have high albedo surfaces coated with fresh ice in the last ~100 Myr. We examine possible resurfacing processes and find none that are plausible. To avoid the influence of cosmic radiation that darkens and reddens most icy surfaces on times scales > ~100 Myr, the family members must be unusually depleted in carbon, or else the collision that created the family occurred so recently that the parent body and fragments have not had time to darken. We also find a rotation period of 4.845 (+/- 0.003) h with amplitude 0.26 (+/- 0.04) mags for 2003 $OP_{32}$.

1. INTRODUCTION

Recently, Brown et al (2007a) presented evidence for a family of six Kuiper Belt Objects (KBOs) with orbits and reflectance spectra similar to its largest member, the Pluto-sized body 2003 $EL_{61}$. All the bodies have neutral to blue spectral reflectance at visible wavelengths, unusually deep water-ice absorption bands in the near IR, and the absence of spectral features from methane or other volatiles (Brown et al 2007a, Barkume et al 2008, Pinilla-Alonso et al 2007). Observations of one of two small satellites orbiting 2003 $EL_{61}$ (Brown et al 2005) and of the rotational light curve of the primary (Rabinowitz et al 2006) show that 2003 $EL_{61}$ is a fast rotating, elongated body with a density exceeding 2600 kg m$^{-3}$ and a visual geometric albedo, $p_v = 0.7$ +/- 0.1, assuming the shape has relaxed to that of a fluid body in hydrostatic equilibrium. Thermal IR measurements with the Spitzer Space Telescope (Stansberry et al 2008) yield $p_v = 0.84$ +/- 0.09, thus confirming the high albedo and supporting the shape model. Given the spectral similarity of the family and the unusual rotation state of the largest member, Brown et al propose that all the smaller family members are fragments originally ejected from the mantle of 2003 $EL_{61}$ in a collision that spun up the primary and created the satellites. From numerical modeling of the orbital evolution of the family, Ragozzine & Brown (2007) estimate the breakup occurred more than 1 Gyr ago.

While the evidence pointing to an ancient collisional breakup of 2003 $EL_{61}$ is strong, this hypothesis presents a new question. Why have the visible colors of the fragments remained conspicuously neutral instead of gradually reddening over time owing to the irradiation of carbon-bearing molecules in the surface ice? This process is thought to be responsible for the red color of most TNOs and to occur over a timescale of 1 Myr to 1 Gyr (Luu & Jewitt 1996, Cruikshank et al 1998, Brunetto et al 2006, Pinilla-Alonso et al 2007). For bodies at least the size of 2003 $EL_{61}$ and with a substantial fraction of surface volatiles (e.g. Eris and Pluto), a neutral color is expected owing to periodic thawing and refreezing of surface ices (Hansen & Paige 1996, Olkin et al 2007, Dumas et al 2007). But for bodies the size of the 2003 $EL_{61}$ collisional retinue (100-300 km), surface gravity is too low for the retention of these volatiles (Schaller & Brown 2007a). There is no atmosphere that can freeze out and hide a reddened surface. Ragozzine and Brown therefore suggest that the original mantle of 2003 $EL_{61}$ may have been depleted in the organic molecules necessary for reddening owing to heating and differentiation at the time of formation. If family members are mantle fragments, then they must also be depleted of organics. The differentiation would also have brought the denser, rocky material to the core of the 2003 $El_{61}$, which is consistent with the observations. Alternatively, Pinilla-Alonso et al (2007) suggest that 2003 $EL_{61}$ might have originated from a population of carbon-depleted bodies in the solar nebula. They also suggest that a process, such as cryovolcanism, might be resurfacing the body with unirradiated ice from its interior on relatively short timescales.

In this paper, we address these scenarios by examining the photometric properties of five of the known family members: 2003 $EL_{61}$, 2002 $TX_{300}$, 2003 $OP_{32}$, 2005 $RR_{43}$ and 1995 $SM_{55}$. We observed these bodies recently as part of our ongoing photometric survey of distant icy bodies with the SMARTS 1.3m telescope at Cerro Tololo. We have already reported our results for 2003 $EL_{61}$, 2002 $TX_{300}$ and other KBOs in earlier publications (Schaefer & Rabinowitz 2002, Rabinowitz et al 2006, and Rabinowitz et al 2007). Here we present new

results for 2003 OP$_{32}$, 2005 RR$_{43}$, and 1995 SM$_{55}$. and also for KBOs (90568) 2004 GV$_9$, (84522) 2002 TC$_{302}$, (145451) 2005 RM$_{43}$, and 2003 AZ$_{84}$, which are not family members. To date, we have measured the colors and phase curves of more than 31 KBOs and Centaurs. All the observations consist of long time series measurements in B-, V-, and I-band filters sufficient to determine the surface colors and the slope of the opposition phase curve (phase coefficient β) at small phase angles, α < 2.0 deg, in each band. The shape of the phase curve depends on the albedo, the porosity and granular structure of the surface, and the surface scattering properties. The phase curve is therefore a sensitive probe of the resurfacing and irradiation processes that are known to alter the surface properties.

Remarkably, we find that all five of the 2003 EL$_{61}$ family members have both extremely neutral color and nearly flat phase curves, a combination we observe for no other bodies in the Kuiper Belt except the icy dwarf planets (Pluto, Eris, 2005 FY$_9$) and only one other KBO (2003 AZ$_{84}$). The remaining KBOs and Centaurs in our survey have significantly steeper phase curves and/or significantly redder colors. The only other known bodies in the outer solar system with photometric properties matching the 2003 EL$_{61}$ family are the bright icy satellites. Comparing the phase curves of all the known icy bodies, including satellites and KBOs, we also find that $p_v$ and β are strongly anti-correlated. Given the low β values we measure for the entire 2003 EL$_{61}$ family, their albedos are probably all high like that of 2003 EL$_{61}$. Interestingly, the surfaces of the dwarf planets and the brightest icy satellites are either suspected or known to be refreshed on 1000 yr to 100 Myr time scales, either by exogenic or endogenic processes. Our results therefore suggest that some process is also resurfacing the 2003 EL$_{61}$ family and preserving their youthful appearance.

II. OBSERVATIONS

The observations we report here for the 2003 EL$_{61}$ family were measured and analyzed as described by Rabinowitz et al (2007). Briefly, all observations are taken by on-site operators at Cerro Tololo using the 1.3-m telescope of the Small and Moderate Aperture Research Telescope System (SMARTS) consortium. Images are recorded with the optical channel of the permanently mounted, dual infrared/optical CCD camera known as A Novel Dual-Imaging Camera (ANDICAM). The optical channel is a Fairchild 2Kx2K CCD that we binned in 2x2 mode to obtain 0.37" per pixel and a 6.3' x 6.3' field of view. All exposures are auto-guided, and typical seeing is 1-2". Because the telescope is queue-scheduled for shared use by all members of the SMARTS consortium, we are able to obtain ~15 minutes of observing time per target every night or every other night for the entire duration of each objects apparition. We typically observe two targets per night, each with a sequence of three or four exposures (B-V-V-I or B-V-I, or V-V-I). Users of the telescope share dome and sky flats, bias frames, and observations in B, V, R, and I of Landolt stars taken on all photometric nights. Table 1 lists the range of observation dates, the number of observations (N$_{obs}$), the minimum and maximum phase angles (α$_{min}$ and α$_{max}$), and the average of the measured V magnitudes (<V>) for each of the targets we discuss in this paper.

For each target we find the solar phase curve, which is the reduced magnitude (the apparent magnitude normalized to 1 AU distance to Earth and the Sun) as a function of α in each filter band. The slope of a linear fit to each phase curve is the phase coefficient in each band ($\beta_B$, $\beta_V$, and $\beta_I$). Table 2 lists the average phase coefficient, $\beta_{avg} = (\beta_B + \beta_V + \beta_I)/3$, for each target. Consistent with our previous analysis of 25 TNO and Centaur phase curves (Rabinowitz et al 2007), we find that each phase curve is linear over the observed range of phase angles and that $\beta$ is not significantly wavelength dependent. We also determine the absolute magnitudes in each band ($B_0$, $V_0$, and $I_0$), which is the reduced magnitude extrapolated to zero phase. Table 2 lists $V_0$ and also the extrapolated opposition colors $B_0$-$V_0$ and $V_0$-$I_0$ for each target. For 2002 $TC_{302}$ we have no B-band observations and we report only the V- and I-band results and their averages. The values listed for 2003 $EL_{61}$ and 2002 $TX_{300}$ are from Rabinowitz et al (2007).

The results listed in Table 2 for 2003 $OP_{32}$ take into account a single-peaked rotation curve with period of 4.845 (+/- 0.003) h and with peak-to-peak amplitude 0.26 (+/- 0.04) mags that we derived from the observations. Figure 1 shows the resulting curve. This we obtained by initially estimating the linear phase dependence of the B, V, and I measurements, subtracting the phase dependence in each band, combining all these corrected observations into a single data set, and then binning the combined set by rotational phase (with bin size equal to 1/20 of the rotation period). We then subtracted this rotational phase curve from the uncorrected observations in each band to determine the rotationally corrected phase coefficients. See Rabinowitz et al (2006, 2007) for a detailed description of the method. We made a similar correction to the phase curve of 2003 $AZ_{84}$ after deriving a rotation curve using the period of 6.71 h reported by Sheppard & Jewitt (2003). Although these corrections did not significantly alter the results, they did significantly reduce the residuals to our linear fits.

Figure 2 shows $\beta$ versus $V_0$-$I_0$ for all the KBOs and Centaurs for which we have reduced observations to date. This includes the new observations we present here and those presented in Rabinowitz et al (2007). The figure also shows results for Pluto using the phase curve reported by Buratti et al (2003) and colors reported by Stern & Yelle (1999). Figure 3a shows the band-averaged phase curve for each of the 2003 $EL_{61}$ family members we have observed. To determine these average curves for each target, we first determined the average values, <B-V> and <V-I>, for the full range in $\alpha$. We then subtracted <B-V> from the reduced B-band magnitudes, added <V-I> to the reduced I-band magnitudes, combined these adjusted data with the reduced V-band magnitudes, and then binned the combined set by solar phase angle. The bin size for each target is the multiple of 0.05 degrees closest to 1/10 the observed range in $\alpha$. The plotted magnitudes are the weighted averaged in each bin and the magnitude errors are those of the weighted mean. The curves have also been arbitrarily shifted in the vertical direction for clarity of presentation. In all cases, we see no significant departure from linear dependence on $\alpha$ over the observed range,

III. RESULTS and DISCUSSION

Examination of Fig. 2 shows that the 2003 $EL_{61}$ family members occupy the extreme lower left of the plot where the colors are bluest and the phase curves are flattest (V-I = 0.6-0.8 mag and $\beta_{avg}$ = 0.0-0.1 mag deg$^{-1}$). The dwarf planets Eris, Pluto, and 2005 $FY_9$ are nearby with similarly flat phase curves but slightly redder colors (V-I = 0.80 - 0.85 mag). Most of the remaining KBOs and all the Centaurs are significantly redder (V-I > 0.9 mag) or have larger phase coefficients ($\beta_{avg}$ > 0.1 mag deg$^{-1}$). Within the 2003 EL$_{61}$ family cluster we find only one non-family interloper (2003 AZ$_{84}$, which we discuss further below). With V-I = 0.69 for the Sun (Hainaut & Delsanti 2002), the family members all have V-I within 0.1 mag of solar color, which is unusual for a KBO. Of the 71 KBOs surveyed by Doressoundiram et al (2005) only 7 objects have V-I < 0.8 mag and 4 of these are known 2003 EL$_{61}$ family members. Of the 151 KBOs and Centaurs with both V and I measurements listed in the MBOSS database compiled by Hainaut & Delsanti (2002), only 13 have V-I < 0.8 mag and 4 of these are family members. Thus the 2003 EL$_{61}$ family is in the 5% tail of the KBO color distribution. Using a computer program to simulate the repeated selection of 5 objects at random from the points plotted in Fig. 1, we find a likelihood of 4.7x10$^{-4}$ that all 5 would cluster by chance in the restricted range occupied by the family. This result further emphasizes the unusual similarity of family members, previously recognized only from their spectral and orbital properties.

In Figs 3a and 3b, we compare the phase curves for the family members with those of the closest known icy analogs, all of which have nearly neutral color (to within ~15%) compared to most KBOs. As shown in Fig 3b, fresh white snow is observed to have a linear phase curve at small angles with nearly zero slope. Triton also has a flat phase curve but only for $\alpha$ > 0.2 deg. At smaller angles there is a narrow opposition surge. Our observations of the 2003 EL$_{61}$ family do not extend to such small phase angles, so we cannot determine if any members have a similarly narrow surge. The other examples are Pluto, Europa, Enceladus, and Rhea, all with nearly flat phase curves at small angles. Like Triton, Enceladus and Rhea also exhibit very sharp surges close to $\alpha$ ~ 0.0 deg (Verbiscer et al 2007) but these observations are not yet published. Our comparisons here show that the 2003 EL$_{61}$ family members could have surfaces with structure similar to any of the large bright icy bodies. Other than the dwarf planets, the bright icy satellites, and 2003 AZ$_{84}$, we know of no other bodies providing as good a photometric match over the observed range of phase angles.

One implication of these results is that the 2003 EL$_{61}$ family members are all likely to have high geometric albedos ($p_v$ > 0.35). This is already suspected given the unusually deep water-ice absorption bands in their spectral reflectance and their neutral visible colors. For example, Licandro et al (2006) fit the spectrum of 2002 TX$_{300}$ equally well assuming a bright or dark surface (90% water ice and $p_v$ = 0.44, or 40% water ice and $p_v$ = 0.13) but the dark model requires an unusually large ice grain size (250 um). Barkume et al (2008) fit the 2003 EL$_{61}$ family assuming ice fractions > 80%. But the solar phase curves we observe constrain the problem further. Below we show that for the icy bodies in the solar system with known albedo and with well-measured phase curves, $p_v$ and $\beta$ are strongly anti-correlated. The bodies with the flattest phase curves at small phase angles have the highest albedos. Karkoschka (2001) has already reported this trend for the Uranian satellites, and suggests that the relation derives from the increasing dominance of coherent backscattering (a small angle effect) over shadow hiding (a large angle effect) with increasing albedo. We show below that this relation holds more generally for all known icy bodies, including satellites of Jupiter, Saturn, and Neptune and KBOs. Since we have no reason to suspect that the 2003 EL$_{61}$ family members are exceptions to this relation, and since they have some of the lowest $\beta$ values we have observed for any icy KBOs, we expect they have very high albedos.

The anti-correlation appears in Fig. 4 where we plot $\beta$ versus $p_v$ for known icy bodies (see also Table 3 for tabulated values). The plot includes all KBOs and planetary satellites with measured $p_v$ and with surfaces known to be rich in water or frozen volatiles (the 2003 $EL_{61}$ fragments do not appear because their albedos have never been measured). The values for $\beta$ are the band-averaged coefficients we have determined from our own observations of KBOs, or else they are from linear fits we make to published phase curves in the range $\alpha = 0.2$-$2.0$ deg. For consistency, we include only those bodies for which the phase curves have been well observed in this narrow $\alpha$ range. Our inclusion of KBOs requires this restriction because they are too distant to be observed at larger phase angles and are not often observable at smaller angles. Furthermore, as we noted above some icy bodies exhibit a non-linear brightness surge at very small angles for which a linear fit is not meaningful. Figure 4 includes 13 of the 14 icy satellites listed in Table IV of the review by Clark et al (1987). We exclude Hyperion because the phase curve is not known at small angles. Note that some of the bright Saturnian satellites (Enceladus, Tethys) have $p_v$ >100% owing to sharp opposition surges at zero phase (Verbiscer et al 2007). The figure also includes all the KBOs and Centaurs observed by Barkume et al (2008) for which they compute a surface ice fraction greater than 20% and for which $\beta$ and $p_v$ are known (2003 $EL_{61}$, Orcus, Quaoar, Ixion, 2003 $AZ_{84}$, and 1999 $TC_{36}$). Also included are satellites Phoebe, Triton, Nereid, and Charon, and the dwarf planets Eris, Pluto, and 2005 $FY_9$. For Callisto, a satellite for which $\beta$ differs significantly on its leading and trailing hemispheres (Thompson & Lockwood 1992), we take the hemispheric average and show error bars spanning the range. Where there are multi-band observations reported for a given target, we fit the phase curve in each band independently and plot the average.

Figure 4 shows that with the exception of 2003 $AZ_{84}$ (labeled "K4" in the figure) the only bodies with $\beta < 0.1$ mag deg$^{-1}$ are those with $p_v > 0.35$. Their median albedo is $p_v = 0.80$. Nearly all the bodies with $\beta > 0.1$ mag deg$^{-1}$ have much lower albedos (median $p_v = 0.23$). For all the data taken together, the Spearman rank-order correlation coefficient is -0.57 with 24 degrees of freedom, which would have a likelihood of 0.12 % if the data were uncorrelated (Press et al 1986). The 2003 $EL_{61}$ family members, all with average phase coefficients < 0.1 mag deg$^{-1}$ and ice-rich surfaces, are thus likely to have $p_v > 0.35$ if they follow the relation (they are likely to fall within the region in Fig. 4 bounded by the dashed lines). It might be argued that some bodies have flat phase curves not because of their high albedos, but because of their porous surfaces (see for example Domingue & Verbiscer 1997). Both porosity and grain size are important factors influencing the width of the opposition surge, and hence the value of $\beta$ (Hapke 1993). If porosity and grain size are the determining properties, then Fig. 4 shows that $p_v$ must be strongly correlated with these properties for icy bodies. However, this need not be the case. In laboratory measurements of soot-chalk mixtures Shkuratov et al (2002) observe an anti-correlation between $p_v$ and $\beta$, similar to the one we find that is not a porosity effect. In any case, we know already that 2003 $EL_{61}$ has a high albedo while Grundy et al (2005) find $p_r > 0.19$ for family member 2002 $TX_{300}$. Hence we find the low $\beta$ values of the 2003 $EL_{61}$ family to be compelling evidence that their albedos are at least as high as Charon's, and perhaps matching those of dwarf planets and the brightest satellites.

To understand the significance of this result, we note here that nearly every icy body in the solar system with an albedo in excess of ~0.4 is known or suspected to be resurfaced on short time scales (<~ 100 Myr) either by endogenic or exogenic processes. For Pluto and Triton, the resurfacing results from repeated sublimation and condensation of their nitrogen atmospheres with each orbit cycle (Grundy and Young 2004, Olkin et al 2007). Similar atmospheric cycles have likely resurfaced Eris (Dumas et al 2007). In the case of the brightest icy moons of Saturn (i.e. Mimas, Enceladus, Tethys, Dione, and Rhea), photometric and spectroscopic observations show that water-ice particles from Saturn's E-ring, whose source is the active plume on Enceladus (Kargel 2006), are resurfacing these bodies (Burrati et al 1990, Filacchione et al 2007, Verbiscer et al 2007). In the case of Chiron, Cook et al (2007) have identified both crystalline water ice and hydrated ammonia on the surface. As both species are short lived under bombardment by solar UV photons and radiation, the authors argue that Chiron's surfaces is replenished on ~100 Myr timescale by cryovolcanism. In the Jovian system there are only two bright moons (Europa and Ganymede), and both are suspected to have internal oceans that have breached their surfaces in the past (Carr et al 1998, McCord et al 2001). Zahnle et al (2003) estimate a young surface age of < 70 Myr for Europa based on crater counts, implying recent geologic activity. Furthermore, ongoing sputtering by plasma carried by Jupiter's magnetic field is though to redistribute water frost on the surfaces of both Europa and Ganymede, creating polar and hemispheric differences in their albedo and color (Tiscareno & Geissler 2003, Khurana et al 2007). Thus, both exogenic and endogenic process may be responsible for refreshing the surfaces of the icy Jovian moons.

In contrast to the bright icy bodies, there is no expectation that dark icy bodies have recently refreshed surfaces. Saturn's darkest icy satellite, Phoebe, orbits too far from the planet to be affected by plasma or ring particles. The dark icy KBOs represented in Fig. 4 (Quaoar, Orcus, Ixion, 1999 $TC_{36}$, 2003 $AZ_{84}$) are too small (<1000 km, see Stansberry et al 2008) to retain substantial amounts of volatiles on their surfaces over the age of the solar system (Schaller & Brown 2007a). For these bodies, there is no expectation of recent resurfacing by cycles of atmospheric freeze out. For example, Schaller & Brown (2007b) observe no $N_2$ or CO on Quaoar, and only small amounts of methane and/or ethane (~10% fraction). They consider Quaoar a transition object, with a marginal volatile fraction surviving long-term sublimation. Earlier observations by Jewitt and Luu (2004) of hydrated ammonia and crystalline water ice on Quaoar's surface led them to propose recent cryovolcanic activity, but the presence of ammonia hydrate appears to be discounted by Schaller & Brown. Similarly, the dark icy satellites of Uranus have no atmospheres or surface volatiles subject to cycles of sublimation and freeze out. Zhanle et al (2003) estimate their geologic surface ages > 1 Gyr from crater counts.

In summary, the close resemblance of the 2003 $EL_{61}$ family members to the brightest icy bodies in the solar system leads us to suspect they have young surface ages. In all observable respects – visible color, near-IR reflectance, light scattering properties – they appear to have surfaces like Charon or other bright icy satellites. Given that all the bright icy bodies observed in the solar system have been resurfaced in the past 100 Myr by various processes that provide them with their unique reflectance properties, we have good reason to expect that the 2003 $EL_{61}$ family members also have high albedos and have been resurfaced in the same time period. The observations by Barkume et al (2008) that surface ice on the family members is crystalline supports this result because crystalline water ice is expected to convert to amorphous ice on a timescale ~10 Myr owing to cosmic radiation (Jewitt & Luu 2004). The resurfacing process itself remains open to question. Except for 2003 $EL_{61}$, the bodies are too small for cryovolcanic activity and too small to retain a volatile atmosphere that could periodically freeze out. Perhaps micrometeorite bombardment has brightened their surfaces, exposing fresh ices below a dark surface, but then it is not clear why this same process has not brightened any non-family KBOs of comparable size. If the surfaces of the family members have always been devoid of organic material, then it is conceivable that they could have escaped the effects of irradiation that have darkened and reddened the ancient surfaces of other KBOs. But there are no other known examples of such carbon-depleted bodies in the Kuiper Belt, even though collisional disruptions of other KBOs like 2003 $EL_{61}$ should have been frequent events early in the solar system's history. Furthermore, the population of impacting meteorites in the Kuiper Belt should bring carbon to an initially pure water-ice surface over time. A fresh veneer of water ice frost, as is commonly observed for the bright icy satellites, provides a more natural explanation for the apparent carbon depletion of the 2003 $EL_{61}$ family. On the other hand, if the fragmentation event that created the family occurred much more recently than estimated, this could also explain the family's youthful appearance. Such a scenario, however, would be extremely unlikely given the low population density of the Kuiper Belt (Durda & Stern 2000).

This work was supported by NASA under Planetary Astronomy grant No. NNX07AK71G and NNX07AK67G. Special thanks to Jenica Nelan and Michelle Buxton for scripting the SMARTS observations.

FIGURE CAPTIONS

Figure 1. Brightness versus rotational phase with period 4.845 h for 2003 OP$_{32}$. This was obtained by subtracting the phase angle dependence from the solar phase curve, determining the rotational phase of each observation, and evaluating the average magnitude for each phase bin of width 1/20 of a full cycle.

Figure 2. Average phase coefficient (β) versus V-I for the 2003 EL$_{61}$ family (large grey squares), dwarf planets Eris, 2005 FY$_9$, and Pluto (white circles), Centaurs (dark grey diamonds), and for TNOs (small black squares). Data for Pluto are from Buratti et al (2003) and Stern & Yelle (1999). This shows the strong correlation between β and V-I, with bodies of neutral color having the flattest phase curves.

Figure 3. (a) Band-averaged reduced magnitude versus solar phase angle for the 2003 El61 family members 2003 EL$_{61}$, 2002 TX$_{300}$, 2003 OP$_{32}$, 2005 RR$_{43}$, and 1995 SM$_{55}$; (b) reduced magnitude versus solar phase angle for bright icy analogs: fresh snow (0.63- μm data from Shuratov et al 2002), Triton (average of R-band data from Buratti et al 2007), Pluto (B-band data from Tholen & Tedesco 1994), Europa (leading hemisphere at 0.47 μm from Domingue et al 1991), Enceladus (0.44-μm data from Verbiscer et al 2005), and Rhea (leading hemisphere at 0.47 μm from Domingue, Lockwood, & Thompson 1995). Over the range of phase angles for which the 2003 EL61 family has been observed, the bright icy analogs have similary flat and linear phase curves, suggesting a similar surface composition.

Figure 4. Phase coefficient (β) at small phase angles (α = 0.2 - 2.0 deg) versus geometric albedo (p$_v$) for bodies showing significant ice absorption features in their reflectance spectra: KBOs (grey squares), satellites (white circles) and dwarf planets (grey circles). This shows a significant anti-correlation between β and p$_v$ for the known icy bodies in the solar system. The box delimited by the dashed lines shows the range of β values observed for the 2003 EL$_{61}$ family. This box also delimits the likely p$_v$ range for the family given the location in the distribution of other icy bodies with similar β values. See Table 3 for references and label explanations.

Table 1. Observing Circumstances.

| Name | $\langle V \rangle$ | $\alpha_{min}$ | $\alpha_{max}$ | $N_{obs}$ | Observing Dates |
|---|---|---|---|---|---|
| 2003 OP$_{32}$ | 20.3 | 0.44 | 1.38 | 78 | 2006 Jul 17 - 2006 Nov 23 |
| 2005 RR$_{43}$ | 20.0 | 0.45 | 1.47 | 144 | 2006 Oct 17 - 2008 Feb 25 |
| 1995 SM$_{55}$ | 20.5 | 0.25 | 1.45 | 60 | 2007 Oct 10 - 2008 Feb 02 |
| 2003 AZ$_{84}$ | 20.4 | 0.19 | 1.11 | 67 | 2007 Jan 09 - 2008 Jan 07 |
| 2002 TC$_{302}$ | 21.0 | 0.13 | 1.16 | 97 | 2006 Sep 28 - 2007 Jan 07 |
| 2004 GV$_9$ | 20.3 | 0.05 | 1.49 | 413 | 2005 Feb 15 - 2006 Aug 23 |
| 2005 RM$_{43}$ | 20.1 | 0.48 | 1.61 | 142 | 2006 Oct 16 - 2007 Mar 03 |

Table 2. Measured Colors, Phase Coefficients, and Reduced $\chi^2$.

| Name | $V_0$ (mag) | $V_0$ error | $B_0-V_0$ (mag) | $B_0-V_0$ error | $V_0-I_0$ (mag) | $V_0-I_0$ error | $\beta_{avg}$ (mag deg$^{-1}$) | $\beta_{avg}$ error |
|---|---|---|---|---|---|---|---|---|
| 2003 EL$_{61}$[a] | 0.428 | 0.011 | 0.646 | 0.015 | 0.686 | 0.014 | 0.097 | 0.007 |
| 2002 TX$_{300}$[a] | 3.365 | 0.044 | 0.869 | 0.053 | 0.636 | 0.067 | 0.076 | 0.029 |
| 2003 OP$_{32}$[a] | 4.097 | 0.033 | 0.698 | 0.052 | 0.773 | 0.052 | 0.040 | 0.022 |
| 2005 RR$_{43}$[a] | 4.125 | 0.071 | 0.790 | 0.076 | 0.693 | 0.076 | 0.010 | 0.016 |
| 1995 SM$_{55}$[a] | 4.490 | 0.030 | 0.628 | 0.023 | 0.683 | 0.050 | 0.060 | 0.027 |
| 2003 AZ$_{84}$ | 3.775 | 0.019 | 0.615 | 0.030 | 0.649 | 0.036 | 0.061 | 0.026 |
| 2002 TC$_{302}$ | 4.225 | 0.034 | … | … | 1.319 | 0.056 | 0.118 | 0.040 |
| 2004 GV$_9$ | 4.263 | 0.020 | 0.843 | 0.028 | 1.150 | 0.026 | 0.127 | 0.011 |
| 2005 RM$_{43}$ | 4.468 | 0.019 | 0.590 | 0.038 | 0.726 | 0.036 | 0.179 | 0.018 |

[a] 2003 EL$_{61}$ family

Table 3. Measured albedo and phase coefficients for icy bodies.

| Name | Label[a] | $p_v$ (%) | $p_v$ error (%) | $\beta$[b] (mag deg$^{-1}$) | $\beta$ error (mag deg$^{-1}$) | Ref.[c] |
|---|---|---|---|---|---|---|
| Europa | J2 | 87.5 | 11.5 | 0.046 | 0.006 | 6, 6 |
| Ganymede | J3 | 44.5 | 1.5 | 0.042 | 0.016 | 14, 7 |
| Callisto | J4 | 21.5 | 1 | 0.104 | 0.021 | 14, 8 |
| Enceladus | S2 | 137.5 | 0.8 | 0.070 | 0.007 | 19, 11 |
| Tethys | S3 | 122.9 | 0.5 | 0.020 | 0.010 | 19, 9 |
| Dione | S4 | 99.8 | 0.4 | 0.010 | 0.010 | 19, 9 |
| Rhea | S5 | 94.9 | 0.3 | 0.066 | 0.009 | 19, 13 |
| Iapetus | S8 | 45 | 2 | 0.065 | 0.016 | 1, 10 |
| Phoebe | S9 | 6 | 2 | 0.180 | 0.035 | 1, 12 |
| Ariel | U1 | 48.9 | 2.1 | 0.246 | 0.028 | 15, 15 |
| Umbriel | U2 | 25.1 | 1.1 | 0.229 | 0.029 | 15, 15 |
| Titania | U3 | 32.8 | 1.4 | 0.214 | 0.008 | 15, 15 |
| Oberon | U4 | 30.9 | 1.3 | 0.219 | 0.031 | 15, 15 |
| Miranda | U5 | 44.9 | 2.2 | 0.244 | 0.041 | 15, 15 |
| Triton | N1 | 82 | 2 | -0.001 | 0.001 | 2, 16 |
| Nereid | N2 | 28.5 | 2 | 0.168 | 0.032 | 3, 17 |
| Charon | P1 | 37.5 | 1.5 | 0.087 | 0.008 | 4, 4 |
| Eris | D1 | 69 | 11 | 0.079 | 0.033 | 5, 18 |
| Pluto | D2 | 58 | 8 | 0.034 | 0.006 | 4, 4 |
| 2005 FY$_9$ | D3 | 78 | 9 | 0.060 | 0.018 | 5, 18 |
| 2003 EL$_{61}$ | D4 | 84 | 9 | 0.097 | 0.012 | 5, 18 |
| Orcus | K1 | 20 | 3 | 0.140 | 0.030 | 5, 18 |
| Quaoar | K2 | 20 | 10 | 0.168 | 0.030 | 5, 18 |
| Ixion | K3 | 17 | 8 | 0.139 | 0.055 | 5, 18 |
| 2003 AZ$_{84}$ | K4 | 12 | 3 | 0.061 | 0.044 | 5, 18 |
| 1999 TC$_{36}$ | K5 | 7.2 | 1 | 0.207 | 0.050 | 5, 18 |

[a] The labels refer to Figure 3.

[b] The phase coefficients are the slopes of linear fits to the observed solar phase curves in the range $\alpha$ = 0.2-2.0 deg. Where multi-band visual data are available, phase coefficients are separately fit to each band and then averaged.

[c] The first reference is for the albedo, the second is for phase curve data.

References: (1) Burns 1986, (2) Pascu et al 2006, (3) Brown et al 1986, (4) Buie et al 1997, (5) Stansberry et al 2008, (6) Domingue et al 1991, (7) Millis & Thompson 1975, (8) Thompson & Lockwood 1992, (9) Noland et al 1974, (10) Franklin & Cook 1974, (11) Verbiscer et al 2005, (12) Kruse et al 1986, (13) Domingue et al 1995, (14) Buratti (1991), (15) Karkoschka 2001, (16) Buratti et al 2007, (17) Schaefer & Tourtellotte 2001, (18) Rabinowitz et al 2007, (19) Verbiscer et al 2007

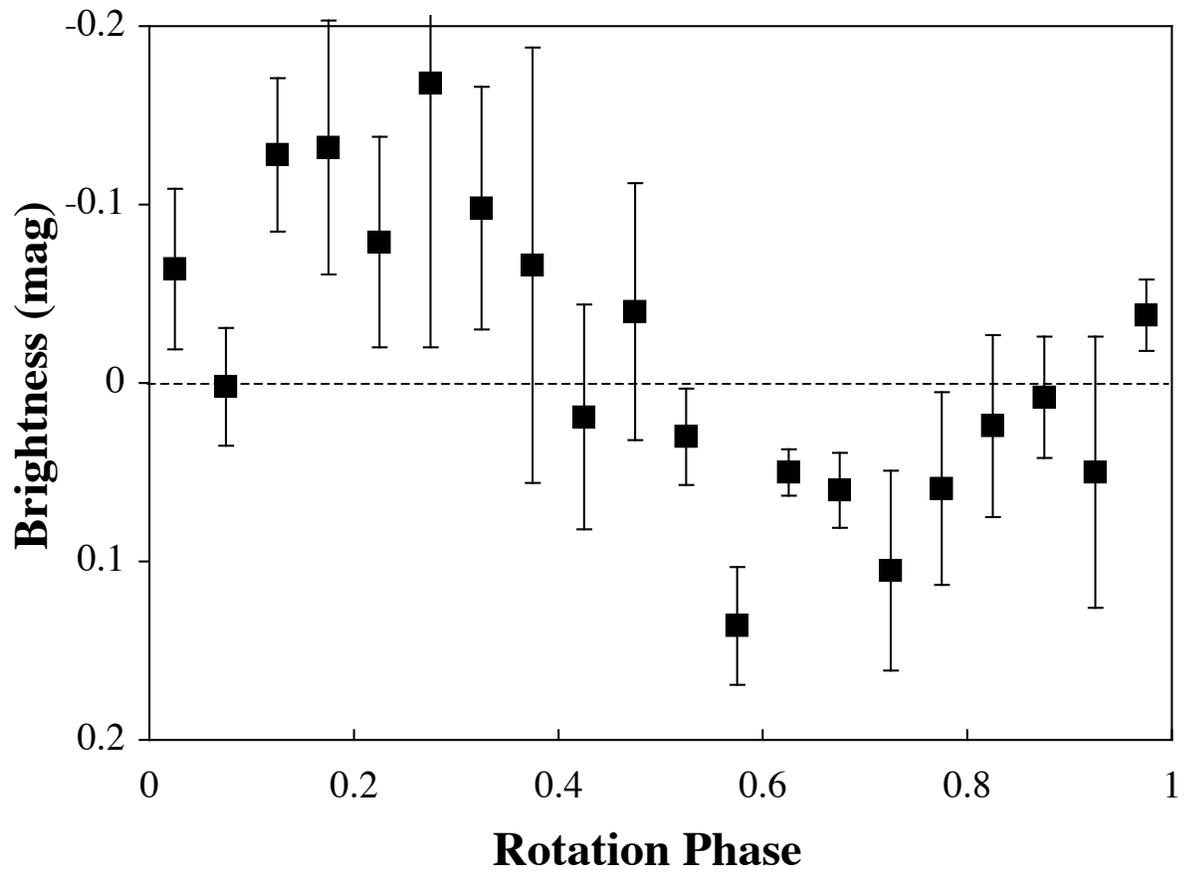

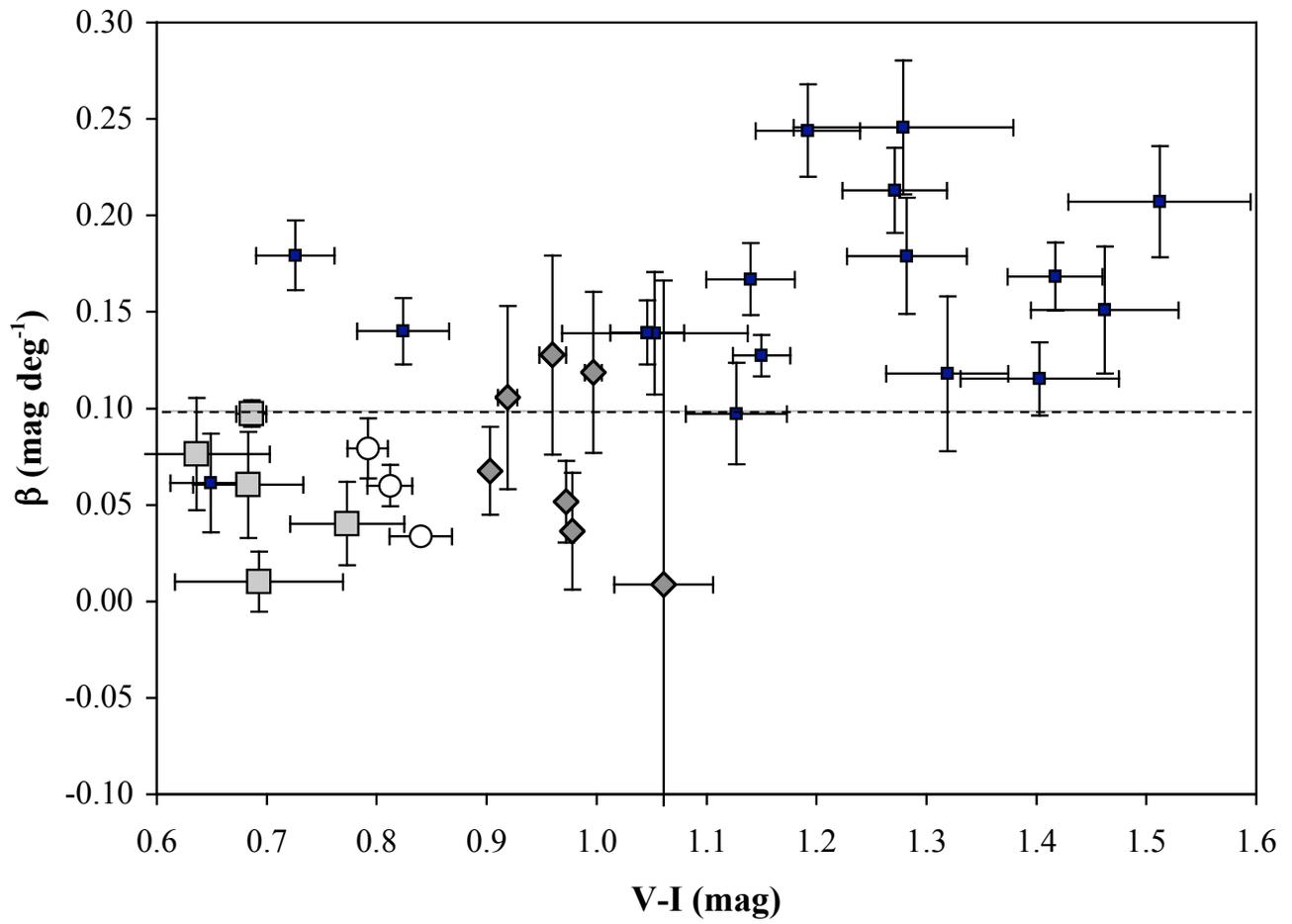

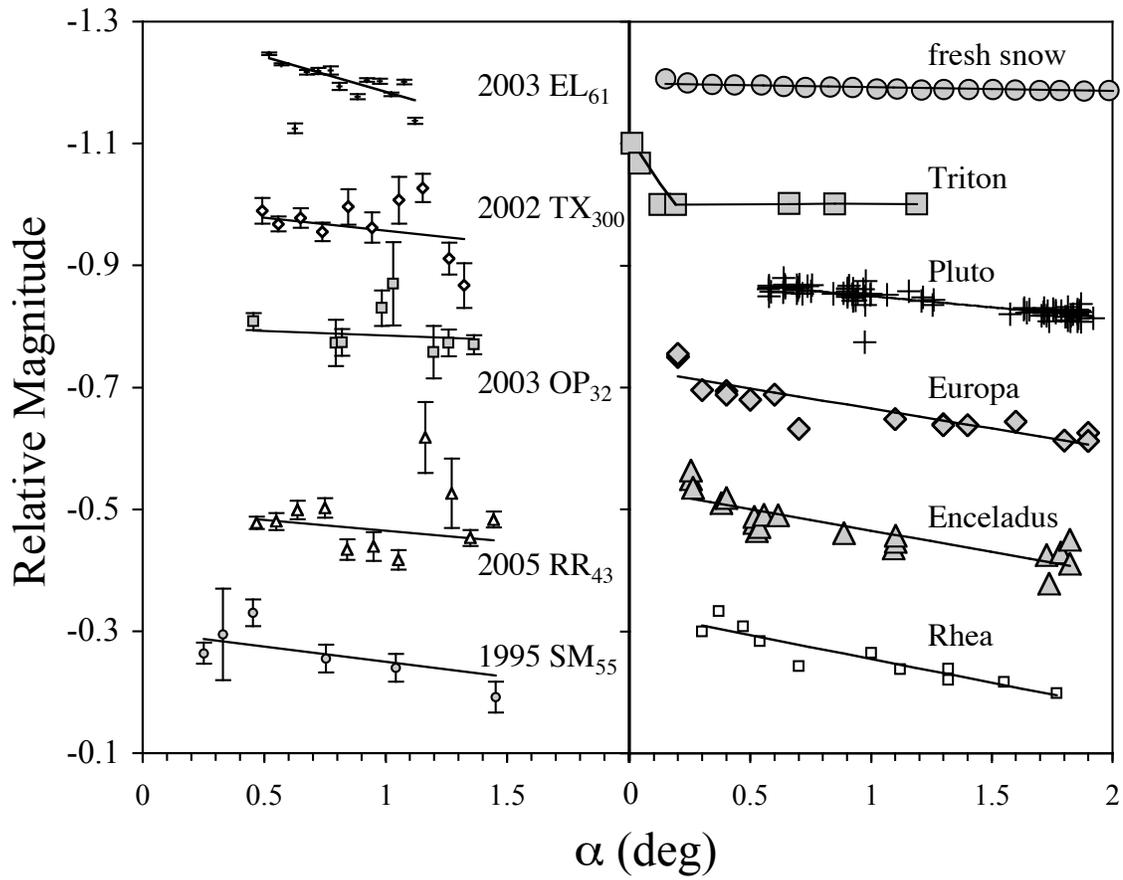

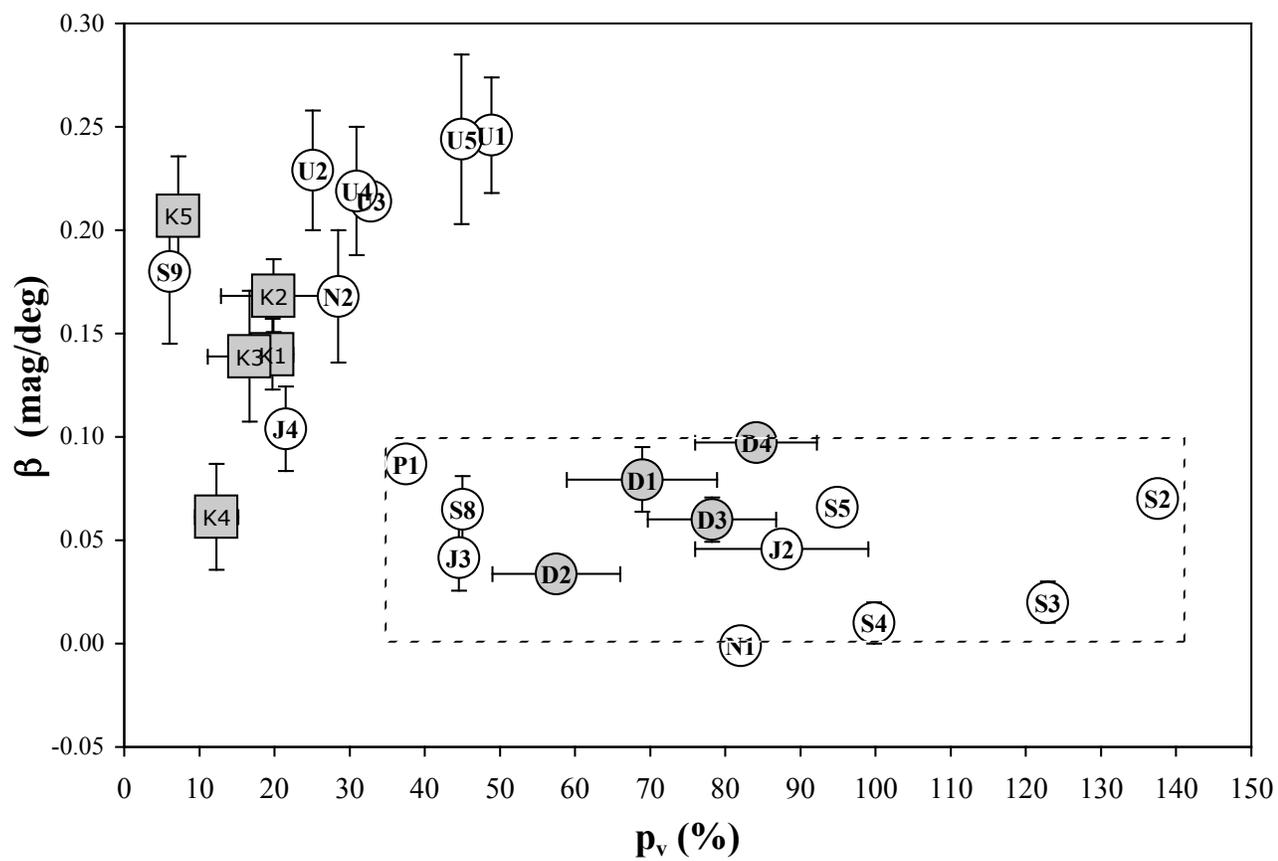